\documentstyle[12pt,epsfig]{article}
\catcode`\@=11
\long\def\@makefntext#1{
\protect\noindent \hbox to 3.2pt {\hskip-.9pt
$^{{\ninerm\@thefnmark}}$\hfil}#1\hfill}                

\def\@makefnmark{\hbox to 0pt{$^{\@thefnmark}$\hss}}  

\def\ps@myheadings{\let\@mkboth\@gobbletwo
\def\@oddhead{\hbox{}
\rightmark\hfil\ninerm\thepage}
\def\@oddfoot{}\def\@evenhead{\ninerm\thepage\hfil
\leftmark\hbox{}}\def\@evenfoot{}
\def\sectionmark##1{}\def\subsectionmark##1{}}

\renewcommand{\thefootnote}{\fnsymbol{footnote}}

\def\sectionc{\@startsection {section}{1}{\z@}{-3.5ex plus -1ex minus
    -.2ex}{2.3ex plus .2ex}{\bf }}
\def\subsectionc{\@startsection{subsection}{2}{\z@}{-3.25ex plus -1ex minus
   -.2ex}{1.5ex plus .2ex}{\it }}
\renewcommand{\section}[1]{\sectionc{#1}\hspace*{\parindent}}
\renewcommand{\subsection}[1]{\subsectionc{#1}\hspace*{\parindent}}
\newcounter{appendixc}
\newcounter{subappendixc}[appendixc]
\newcounter{subsubappendixc}[subappendixc]

\renewcommand{\appendix}[1] {\vspace*{0.6cm}
        \refstepcounter{appendixc}
        \setcounter{figure}{0}
        \setcounter{table}{0}
        \setcounter{equation}{0}
        \renewcommand{\thefigure}{\Alph{appendixc}.\arabic{figure}}
        \renewcommand{\thetable}{\Alph{appendixc}.\arabic{table}}
        \renewcommand{\theappendixc}{\Alph{appendixc}}
        \renewcommand{\theequation}{\Alph{appendixc}.\arabic{equation}}
        \noindent{\bf Appendix \theappendixc #1}\par\vspace*{0.4cm}}

\def\abstracts#1{{

\centering{\begin{minipage}{13.2truecm}\footnotesize\baselineskip=13pt\noindent
        \parindent=0pt #1
        \end{minipage}}\par}}

\renewenvironment{thebibliography}[1]
        {\begin{list}{\arabic{enumi}.}
        {\usecounter{enumi}\setlength{\parsep}{0pt}
\setlength{\leftmargin 0.75cm}{\rightmargin 0pt}
         \setlength{\itemsep}{0pt} \settowidth
        {\labelwidth}{#1.}\sloppy}}{\end{list}}

\newcounter{itemlistc}
\newcounter{romanlistc}
\newcounter{alphlistc}
\newcounter{arabiclistc}

\newcommand{\fcaption}[1]{
        \refstepcounter{figure}
        \setbox\@tempboxa = \hbox{\footnotesize Figure~\thefigure. #1}
        \ifdim \wd\@tempboxa > 6in
           {\begin{center}
        \parbox{6in}{\footnotesize\baselineskip=13pt Figure~\thefigure. #1}
            \end{center}}
        \else
             {\begin{center}
             {\footnotesize Figure~\thefigure. #1}
              \end{center}}
        \fi}

\newcommand{\tcaption}[1]{
        \refstepcounter{table}
        \setbox\@tempboxa = \hbox{\footnotesize Table~\thetable. #1}
        \ifdim \wd\@tempboxa > 6in
           {\begin{center}
        \parbox{6in}{\footnotesize\baselineskip=13pt Table~\thetable. #1}
            \end{center}}
        \else
             {\begin{center}
             {\footnotesize Table~\thetable. #1}
              \end{center}}
        \fi}

\def\@citex[#1]#2{\if@filesw\immediate\write\@auxout
        {\string\citation{#2}}\fi
\def\@citea{}\@cite{\@for\@citeb:=#2\do
        {\@citea\def\@citea{,}\@ifundefined
        {b@\@citeb}{{\bf ?}\@warning
        {Citation `\@citeb' on page \thepage \space undefined}}
        {\csname b@\@citeb\endcsname}}}{#1}}

\newif\if@cghi
\def\cite{\@cghitrue\@ifnextchar [{\@tempswatrue
        \@citex}{\@tempswafalse\@citex[]}}
\def\citelow{\@cghifalse\@ifnextchar [{\@tempswatrue
        \@citex}{\@tempswafalse\@citex[]}}
\def\@cite#1#2{{$\null^{#1}$\if@tempswa\typeout
        {IJCGA warning: optional citation argument
        ignored: `#2'} \fi}}

 1
 1
 1

\font\ninerm=cmr9

\begin{document}

\newcommand{\slt}{\!\!\!/}
\newcommand{\sld}{\!\!/}

\centerline{\normalsize\bf KAON PHOTOPRODUCTION ON THE NUCLEON : }
\baselineskip=15pt
\centerline{\normalsize\bf STATUS AND FUTURE PROSPECTS}

\vspace*{0.6cm}
\centerline{\footnotesize C. BENNHOLD}
\baselineskip=13pt
\centerline{\footnotesize\it Center of Nuclear Studies, Department of Physics,
  The George Washington University,}
\baselineskip=13pt
\centerline{\footnotesize\it Washington, D.C. 20052, USA}
\centerline{\footnotesize E-mail: bennhold@gwis2.circ.gwu.edu}
\vspace*{0.3cm}
\centerline{\footnotesize and}
\vspace*{0.3cm}
\centerline{\footnotesize T. MART, D. KUSNO}
\baselineskip=13pt
\centerline{\footnotesize\it Jurusan Fisika, FMIPA, Universitas Indonesia,}
\baselineskip=13pt
\centerline{\footnotesize\it Depok 16424, Indonesia}
\centerline{\footnotesize E-mail: mart@makara.cso.ui.ac.id}
\vspace*{0.3cm}

\vspace*{0.6cm}
\abstracts{Kaon photo- and electroproduction off the nucleon is investigated
  in the framework of an isobaric model up to energies of about 1 GeV above
  threshold for all six isospin channels. The hadronic coupling constants in
  the model are determined by fitting to the available experimental
  data.  We give a brief status report on the state of these models and
  discuss problems to be attacked in the future.
   We include hadronic form factors and show that they are
   essential for a proper description of the dynamics even though
   their inclusion has to be done carefully in order to
   preserve gauge invariance.
  We also investigate the effects of the $K^0$ electromagnetic form factors
  for two quark models on the longitudinal and transverse differential cross 
  sections. It is found that only the reaction $n(e,e^{\prime }K^0)\Lambda$ 
  is sufficiently sensitive to extract the $K^0$ form factor.}

\normalsize\baselineskip=15pt
\setcounter{footnote}{0}
\renewcommand{\thefootnote}{\alph{footnote}}

\section{Introduction}\label{sec:intro}
For more than three decades, kaon electromagnetic production has been
used as a tool to study strange hadrons and their resonances.  However,
only with the recent advance of high energy, high duty cycle electron
accelerators like TJNAF can the full potential of this reaction be
exploited.  The process
can be performed on nucleons or nuclei.  With the nucleon as a target, the
final state consists of a kaon and a hyperon, and one can study the
reaction mechanism, coupling constants, electromagnetic form factors of the
kaon or hyperon, or even form factors at the hadronic vertices
which until now have usually been neglected.  Using
a nucleus as the target, one can study the hypernuclear spectroscopy and
the $YN$ interaction.

Our understanding of the kaon-baryon interaction is much poorer
than our knowledge of the pion-nucleon force, exemplified by the
uncertainty in the kaon-hyperon-nucleon coupling constants. Unlike the
well established pion-nucleon interaction which yields
the pion-nucleon coupling constant $g_{\pi NN}^2/4\pi$ around 14, the
kaon coupling constants extracted from different reactions (from 
hadronic to electromagnetic) vary wildly as shown in Table \ref{tab:cct}.
For most of the past decades there was a serious discrepancy between
values for the
coupling constants extracted from electromagnetic reactions and those
from hadronic processes which tend to be closer to the SU(3) values.
Recent $(\gamma, K)$ studies~\cite{saghai,terry3}, however, have shown
that the electromagnetic values can be brought into better agreement with the
hadronic analyses.

\begin{table}[t]
\begin{center}
\tcaption{The leading hadronic coupling constants from several models.
         Sets VI-VIII come from fits to isobaric photoproduction models. }
\label{tab:cct}
\begin{tabular}{llccc}
\hline\hline\\
Set &Source & ${\displaystyle \frac{g_{K\Lambda N}}{\sqrt{4\pi}}}$&
${\displaystyle \frac{g_{K\Sigma N}}{\sqrt{4\pi}}}$& Reference\\
[2.5ex]
\hline\\
I&SU(3)&$-4.40$ to $-3.0$&$+0.9$ to $+1.3$&\cite{adel2}\\
II&$K$-$N$ scattering&$~|3.53|$&$~|1.53|$&\cite{anto}\\
III&$Y$-$N$ scattering&$-3.86$&$+1.09$&\cite{timmer}\\
IV & $N{\bar N}\rightarrow Y{\bar Y}$ LEAR data & $-3.92$ & - & \cite{timmer}\\
V&QCD sum rules&$-1.96$&$+0.33$&\cite{choe}\\
VI&$p(\gamma,K^+)\Lambda$, $p(e,e'K^+)\Lambda$&$-3.16$&$+0.80$&\cite{saghai}\\
VII&$p(\gamma,K^+)\Lambda$, $p(e,e'K^+)\Lambda$&$-2.38$&$+0.23$&
\cite{williams}\\ \\
VIII&$\displaystyle \!\!\!\! \left. \begin{array}{l}
p(\gamma,K^+)\Lambda , ~p(e,e'K^+)\Lambda,\\
p(\gamma,K^+)\Sigma^0 ,~p(e,e'K^+)\Sigma^0,\\
p(\gamma,K^0)\Sigma^+ ,~n(e,e'K^+)\Sigma^- \end{array} \right\}$&
$-3.74$&$+0.86$&\cite{terry3}\\ \\
\hline\hline
\end{tabular}
\end{center}
\end{table}

Up to now, most analyses have focused only on the 
$p(\gamma, K^+)\Lambda$ and $p(\gamma, K^+)\Sigma^0$ channels.
The lack of experimental data for other isospin channels was a limiting 
factor for almost three decades. Considerable efforts
have been devoted to model the first reaction, for which most cross section
data are available.  This situation, which lead to
an incomplete understanding of the kaon photoproduction process, is
about to change with new data from SAPHIR for the neutral kaon channel
and the imminent start of experiments in Hall B at TJNAF that will
explore all possible isospin channels.


\section{Isobaric Models}

In general, the $T$-matrix for any photoproduction process can be written
in terms of a Bethe-Salpeter equation
\begin{eqnarray}
M ~ = ~V ~+ ~ V~G~T,
\label{eq:tmatrix}
\end{eqnarray} 
where $V$ represents the driving term for the particular photoproduction
process, $G$ is the meson-baryon two-particle propagator, and $T$ is the
hadronic meson-baryon final state interaction. While approaches using
the above description are becoming increasingly successful in the description 
of pion photoproduction the hadronic final state interaction is usually 
neglected in kaon photoproduction models.  Thus, the $T$-matrix is simply
approximated by the driving term which is usually assumed to be given by 
a series of tree-level Feynman diagrams.

\begin{figure}[!t]
\centerline{\epsfxsize=12cm \epsffile{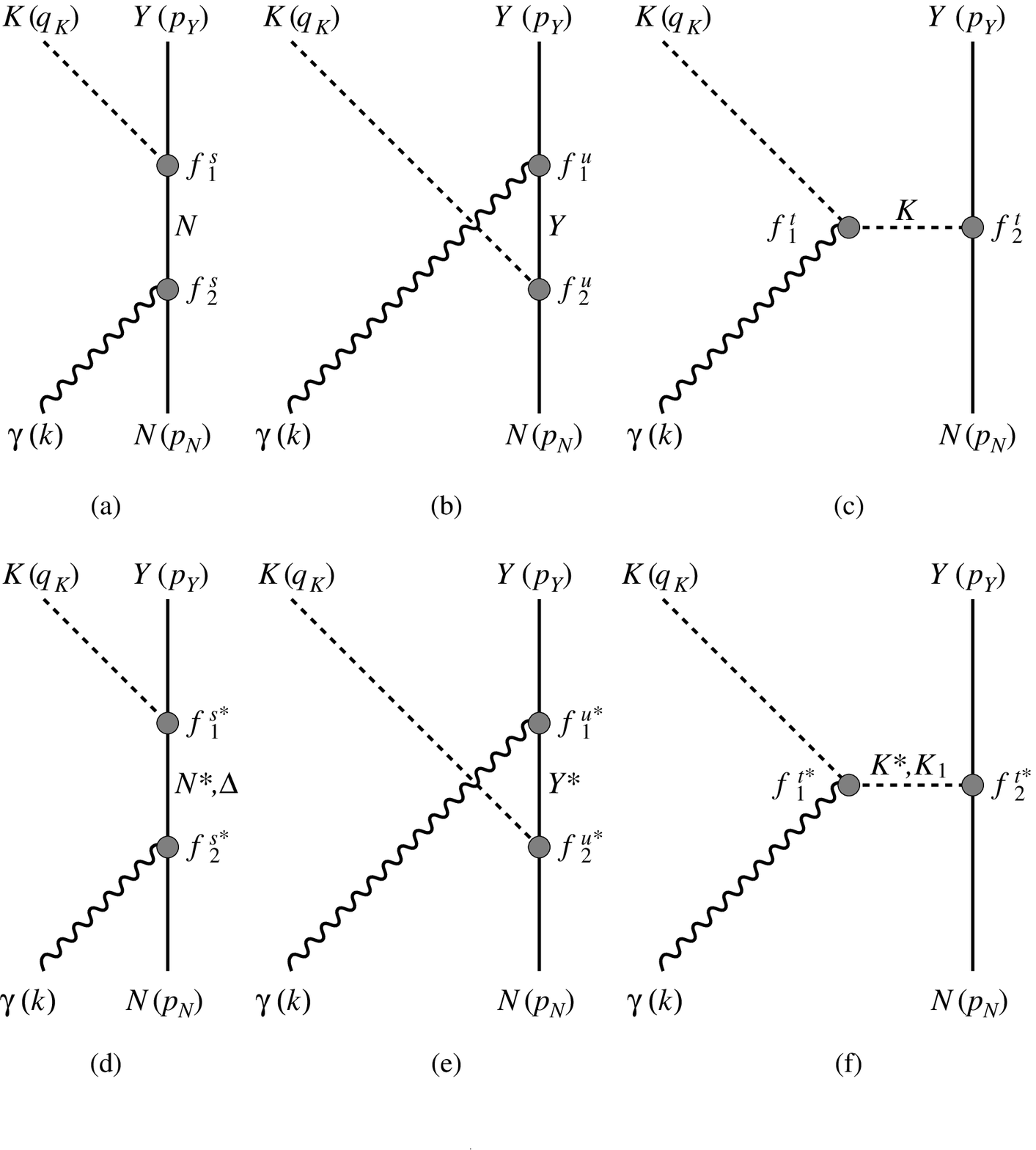}}
\fcaption{Feynman diagrams for kaon photoproduction. The diagrams in (a), (b), 
and (c) represent the $s$-, $u$-, and $t$-channel Born terms, respectively, 
while diagrams (d), (e), and (f) show the corresponding resonance 
contributions. Note that the $\Delta$ resonance contributes only to 
$\Sigma$ production. The vertex factors $f_1$, $f_2$, $f_1^*$, and $f_2^*$, as
well as the corresponding propagators are given e.g. in
Refs.~\protect\cite{abw,terry1}. } 
\label{fig:feynman}
\end{figure}

Figure~\ref{fig:feynman} shows the set of standard 
Feynman diagrams included in the description of the process. 
Besides the Born terms with the nucleon, hyperon and kaon in the respective
$s$-, $u$-, and $t$-channel, $N^*, Y^*$ and $K^*$ resonances are introduced in
the same channels. The coupling constants which are products of strong and 
electromagnetic couplings are then fitted to the available data. Until now 
the quality of the data has neither permitted a clear identification of the 
relevant resonances in the $(\gamma, K)$ process nor allowed a clear
separation of the resonance contributions from the background.  The recent 
study of Ref.~\cite{saghai} gives a very comprehensive description that 
includes resonances up to spin 5/2.  They achieve an excellent fit 
with strong couplings within the SU(3) range.  However, due to the large 
number of resonances the resulting elementary 
operator is rather cumbersome for nuclear applications.

In general, the transition matrix elements can be written as
\begin{eqnarray}
M_{\rm fi} & = & \bar u(p_{Y}) \sum_{j=1}^6 A_{j}(s,t,k^2) M_{j} u(p_{N})~ ,
\label{eq:trans}
\end{eqnarray}
where $s$ and $t$ are the usual Mandelstam variables, and the
the gauge and Lorentz invariant matrices $M_{j}$ are given in many 
references~\cite{saghai,abw,terry1}. The amplitudes $A_{j}$ depend on the
particular model employed for background and resonances.

\subsection{Hadronic form factors and gauge invariance}

No analysis of kaon photoproduction 
has ever included a form factor at the hadronic
vertex.  However, since most of the present isobaric models diverge at 
higher energies, the need for such hadronic form factors has been known
for a long time.  Furthermore, recent work\cite{terry2} on the additional
isospin channels demonstrated that models which give a good description of
the $(\gamma,K^+)$ data can give unrealistically large predictions for the
$(\gamma,K^0)$ channels. As we show below, incorporating a hadronic form 
factor helps alleviate this problem.
On the other hand, it is well known that
the inclusion of form factors at the hadronic vertices, $KYN$, leads to a
problem with gauge invariance, since not every diagram shown in 
Fig.~\ref{fig:feynman} retains gauge invariance by itself. 
The resonance contributions are constructed to be independently gauge 
invariant, but this is not the case for the Born diagrams 
shown in Figs.~\ref{fig:feynman} (a), (b), and (c).  

As an example, for the $p(\gamma,K^+)\Lambda$ channel the amplitude for the
diagram (a) in Fig.~\ref{fig:feynman} is given by 
\begin{eqnarray}
  \label{eq:gi1}
  {\cal M}^p &=& {\bar u}_{\Lambda} \frac{ieg_{K\Lambda N}\gamma_5}{s-m_p^2}~
  \biggl[~\frac{\kappa_p}{m_p}(p_p \cdot k \epsilon \sld - p_p \cdot \epsilon
  k\slt) - (1+\kappa_p) \epsilon \sld k \slt  \nonumber\\
 &&-~\frac{4}{t-m_K^2}~p_p \cdot \epsilon~q_K\cdot k~ \biggr]~u_p~,
\end{eqnarray}
where the last term is not gauge invariant by itself. 
The amplitude from diagram (b) is 
\begin{eqnarray}
  \label{eq:gi2}
  {\cal M}^{K^+} &=& {\bar u}_{\Lambda} \frac{ieg_{K\Lambda N}\gamma_5}{t-
  m_K^2}~\biggl(~\frac{4}{s-m_p^2}~q_K\cdot \epsilon~ p_p\cdot k ~\biggr)~
  u_p~.
\end{eqnarray}
Assuming a point-like hadronic vertex, the term of 
Eq.~(\ref{eq:gi1}) along with Eq. (\ref{eq:gi2}) can be combined together 
to form a gauge-invariant term. 

However, if we include hadronic 
form factors $F_{\rm h}(p_p^2,q_K^2,p_{\Lambda}^2)$ 
at both vertices, those terms will become
\begin{eqnarray}
  \label{eq:ngi}
  \Delta{\cal M} &=& {\bar u}_{\Lambda} \frac{ieg_{K\Lambda N}\gamma_5}{s-
  m_p^2}~\frac{4}{t-m_K^2}~\biggl[~q_K\cdot \epsilon~p_p\cdot k~ F_{\rm h}(
  m_p^2,t,m_{\Lambda}^2) \nonumber\\
  &&-~ p_p\cdot \epsilon~q_K\cdot k~  F_{\rm h}(s,m_K^2,
  m_{\Lambda}^2)~\biggr]~u_p~,
\end{eqnarray}
which will not vanish when substituting $\epsilon\rightarrow k$, 
since the hadronic form factors, 
$F_{\rm h}(m_p^2,t,m_{\Lambda}^2)$ and $F_{\rm h}(s,m_K^2,m_{\Lambda}^2)$, 
have different 
arguments\cite{naus1}.  In order to restore gauge invariance,  
Ohta\cite{ohta} has derived an additional amplitude by making use of
the minimal-substitution prescription. In our formalism, Ohta's additional
amplitude has the form
\begin{eqnarray}
  \label{eq:ohtas}
  \Delta{\cal M}^{\rm Ohta} &=& {\bar u}_{\Lambda} \frac{ieg_{K\Lambda 
  N}\gamma_5}{s-m_p^2}~\frac{4}{t-m_K^2}~\biggl[~q_K\cdot \epsilon~p_p\cdot k~ 
  \left\{ F_{\rm h}(m_p^2,m_K^2,m_{\Lambda}^2)-F_{\rm h}(m_p^2,t,
  m_{\Lambda}^2)\right\}
  \nonumber\\
  &&-~p_p\cdot\epsilon~q_K\cdot k~\left\{ F_{\rm h}(m_p^2,m_K^2,m_{\Lambda}^2)-
  F_{\rm h}(s,m_K^2,m_{\Lambda}^2)\right\}~\biggr]~u_p~,
\end{eqnarray}
where $F_{\rm h}(m_p^2,m_K^2,m_{\Lambda}^2)=1$, i.e. 
if all particles are on-shell. Adding this amplitude to Eq.~(\ref{eq:ngi})
restores gauge invariance, but at the price of removing the off-shell 
hadronic form factors at tree level for non gauge-invariant diagrams.  

A consistent procedure for the inclusion of hadronic form factors
in photoproduction is, however, somewhat complicated, since the form factor
should be different for different channels\cite{gross1}. Such an 
investigation is still underway. For a qualitative study of the influence 
of hadronic form factors on kaon production, we choose
to multiply the whole amplitude with a simple monopole form factor
\begin{eqnarray}
  F_{\rm h}(\Lambda_{\rm c},t) &=& \frac{\Lambda_{\rm c}^2-m_{K}^2}{
  \Lambda_{\rm c}^2-t} ~,
\label{eq:hadff}
\end{eqnarray}
with $\Lambda_{\rm c}$ as a cut-off parameter to be 
determined by our fit. 

\subsection{Results}

\begin{table}[t]
\begin{center}
\tcaption{The coupling constants of set I were obtained by fitting all
         differential and total cross section data except those of charged
         $\Sigma$ and the new Bonn polarization data,  
         set II by including all data.}
\begin{tabular}{|l|r|r|r|r|r|}
\hline\hline
\multicolumn{1}{|c|}{Set}&\multicolumn{2}{|c|}{I}  & \multicolumn{2}{|c|}{II}\\
\hline
\multicolumn{1}{|c|}{~~Coupling Constant~~} & ~$K\Sigma$~~~& 
~$K\Lambda$~~~& ~$K\Sigma$~~~& ~$K\Lambda$~~~\\ 
\hline\hline
&\multicolumn{2}{|c|}{}&\multicolumn{2}{|c|}{}\\
$g_{K\Sigma N}/\sqrt{4\pi}$& \multicolumn{2}{|c|}{~~0.769}& 
\multicolumn{2}{|c|}{~~1.224}\\
$g_{K\Lambda N}/\sqrt{4\pi}$&\multicolumn{2}{|c|}{$-1.911$}&
\multicolumn{2}{|c|}{$-3.092$}\\ [1ex]
\hline
$G_{V}(K^{*})/4\pi$ &0.084&$-0.132$&$-0.080$&$-0.189$ \\
$G_{T}(K^{*})/4\pi$ &$-0.240$&$0.141$&$-0.080$&$-0.123$ \\
$G_{N4}(1650)/\sqrt{4\pi}$&$-0.142$&$-0.033$&$-0.007$&$-0.064$ \\
$G_{N6}(1710)/\sqrt{4\pi}$ &$-0.768$&$-0.133$&2.102&$-0.065$\\
$G_{\Delta (1/2)}(1900)/\sqrt{4\pi}$ &0.120&-&0.234&-\\
$G_{\Delta (1/2)}(1910)/\sqrt{4\pi}$ &0.746&-&$-0.990$&-\\
\hline
$\Lambda_{\rm c}$&\multicolumn{2}{|c|}{-}&\multicolumn{2}{|c|}{0.853}\\
$N$&\multicolumn{2}{|c|}{723}&\multicolumn{2}{|c|}{748}\\
$\chi^2/N$&\multicolumn{2}{|c|}{7.33}&\multicolumn{2}{|c|}{5.98}\\
[1ex]
\hline\hline
\end{tabular}
\label{tab:newcc}
\end{center}
\end{table}

Preliminary results of our fits are displayed in Table \ref{tab:newcc}, where 
all available data in the different isospin channels have been included
in a simultaneous fit. In this
study we have used the complete data set, including preliminary 
Bonn data.  In a previous investigation we found that including the few 
charged $\Sigma$ photoproduction data lead to substantially suppressed
coupling constants $g_{K\Sigma N}$ and $g_{K\Lambda N}$~\cite{terry2}.
This result is changed completely by the inclusion of the hadronic form 
factor.  As shown by sets I and II in Table \ref{tab:newcc},
the hadronic form factor given in Eq.~(\ref{eq:hadff}) 
has a profound influence on the coupling constants
and the $\chi^2$. 
The two leading coupling constants increase to values consistent
with SU(3) (see Table \ref{tab:cct}) and the 
$\chi^2$ is significantly reduced.  We believe that the absence of the 
hadronic form factors in previous models is the main reason for the
disagreement between the hadronic and electromagnetic extractions of the 
kaon-nucleon-hyperon couplings.
Not only the Born couplings but also the resonance couplings are 
significantly changed both in sign and magnitude
when the form factor is included.  The cut-off mass of 850 MeV appears
to be reasonable in comparison to meson-nucleon scattering models.

\begin{figure}[t]
\centerline{\epsfxsize=10cm \epsffile{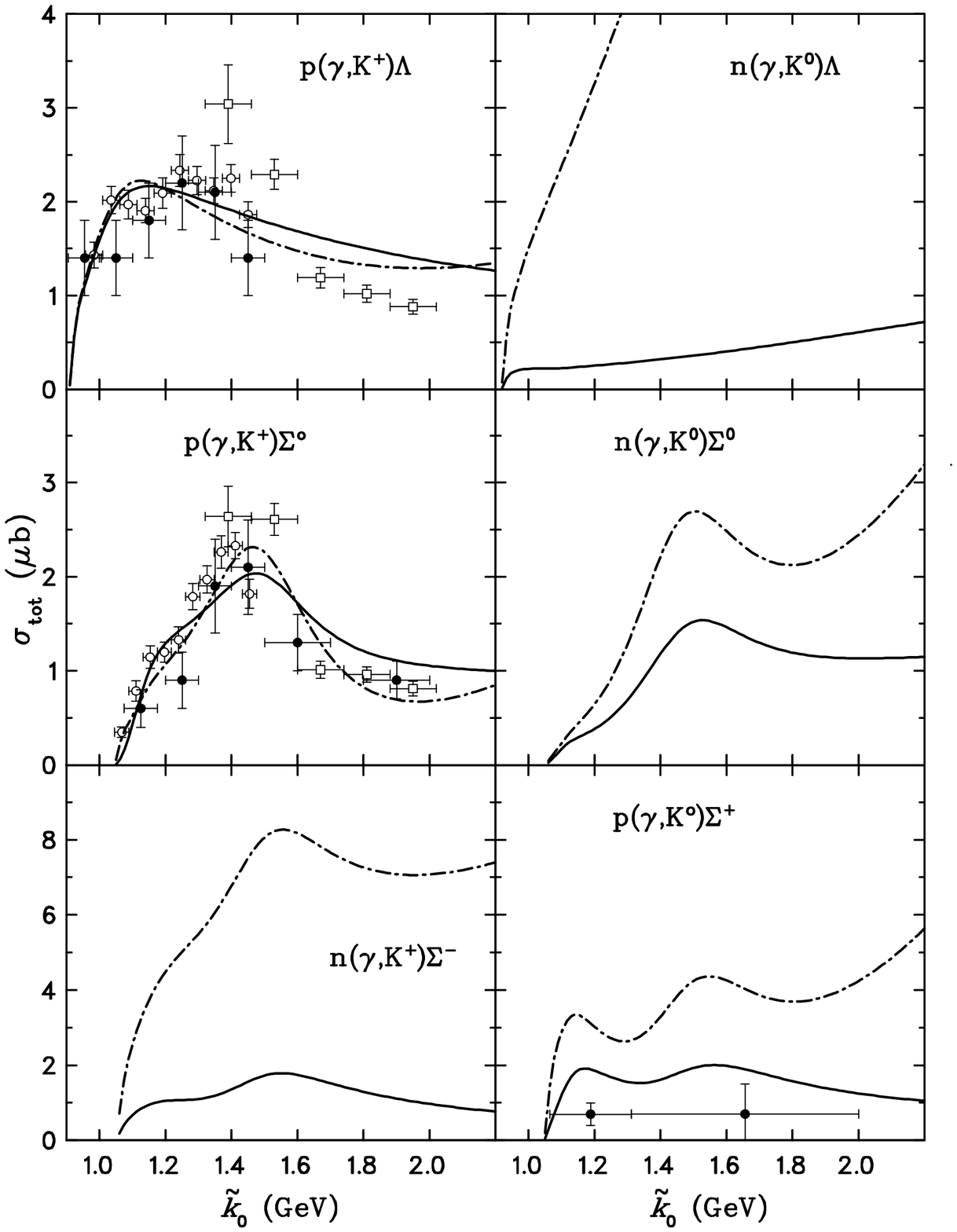}}
\fcaption{Total cross sections  for the six isospin 
         channels of kaon photoproduction calculated from 
         models based on set I (dash-dotted lines)  
         and set II (solid lines).  The 
         solid lines are calculations including the hadronic form 
         factor.}
\label{fig:totsix}
\end{figure}

The impact of the hadronic form factors on the total cross sections in 
all six isospin channels is displayed in Fig.~\ref{fig:totsix}. 
While both the $K^+ \Lambda$ and $K^+ \Sigma^0$ data are fairly well 
reproduced in either model it is the neutral $K^0$ and charged $\Sigma$
production where the form factor makes an important difference.
In $K^+\Sigma^-$ production the form factor can reduce 
the total cross section by more than a factor of four 
which is a significant improvement over the previous
models\cite{terry2}.  In general, the form factor improves the 
convergence of the cross sections at the higher energies where most models
tend to overpredict the data by large amounts.
Nevertheless, more resonances will have to be included in a more
realistic model. Below we give a brief list of additional issues
that will need to be addressed in future models:

\begin{enumerate}

\item {\it Pseudoscalar versus pseudovector coupling}

Most models until now have employed pseudoscalar coupling for the 
kaon-hyperon-nucleon vertex, mostly because previous studies~\cite{ben2}
indicated that using pseudovector coupling would lead to a further 
suppression of the leading Born couplings in the fit to the data. However,
with the inclusion of the hadronic form factor these studies have to be 
repeated since this suppression has vanished. In view of the successes
of chiral symmetry arguments in the SU(2) sector one may want to apply
pseudovector coupling even though the larger mass of the kaon may
make those arguments less valid.

\item  {\it Unitarity and crossing}

Neglecting the final meson-baryon interaction in the full $(\gamma,K)$ 
$T$-matrix automatically leads to a violation of unitarity since flux
that can "leak out" into inelastic channels has not been properly
accounted for.  In principle, enforcing unitarity dynamically requires
solving a system of coupled channels with all possible final states. This
is clearly a daunting task, especially since it would require information
on channels such as $\Lambda(K^+,K^+)\Lambda$ for which no experimental
information is available for obvious reasons. Including unitarity properly
would break crossing symmetry which has been imposed by a number of 
models~\cite{saghai,williams}. Crossing can only be maintained at the tree
level but would have to be given up in a coupled channels model. This becomes
apparent when one compares the intermediate hadronic states of $p(\gamma,
K^+)\Lambda$ with those of $p(K^-,\gamma)\Lambda$. While these two processes
are related via crossing at the tree level, the photoproduction process
proceeds though intermediate states with zero strangeness while the radiative
capture reaction requires $S=-1$.  A first step towards 
the enforcement of unitarity would be the inclusion of phases between the
different resonances which could be fitted to the data. Thus, the data
themselves would determine the amount of final state interaction in a 
phenomenological way.

\item {\it Duality}

The idea of duality comes from high-energy hadron-hadron interactions
where it was shown that including all resonances in the $t$-channel along with
all resonances in the $s$-channel would amount to double counting. At present,
it is not clear how and if this issue would apply to low-energy 
kaon photoproduction. Clearly, the very truncated model space in both $s$- 
and $t$-channel probably minimizes double counting, albeit not in a consistent
fashion. While models have been constructed with no $K^*$ resonances that
allowed a fairly good description of the data, no model has attempted the
reverse, eliminating $s$-channel resonances while including as many $t$-channel
contributions as possible. Such a model may not give a good $\chi^2$ since
it would miss the peaks of the resonances in the cross sections but may give
an adequate average description of the cross section.  Future work will have
to verify this possibility. An important difference to duality in 
hadron-hadron interaction is the presence of gauge invariance which requires
that both $s$- and $t$-channels be present at the Born level.  Thus, the double
counting argument could only apply to the resonance contributions in these
channels.
\end{enumerate}

\section{The $K^0$ Electromagnetic Form Factor}


Among the many applications of the kaon electromagnetic production process is
its potential to access electromagnetic form factors of strange hadrons.
The recent completion of TJNAF has sparked increased interest in 
form factors of baryons and mesons. In particular, for the charged 
pion and kaon experimental proposals have been approved at TJNAF
\cite{mack,baker}, while up to now no study has been  performed
to estimate the effect of the neutral kaon form factor on 
experimental observables accessible at this facility. The underlying
reason, as pointed out by Ref.~\cite{buck}, is the difficulty to perform 
a direct measurement of the form factor due to the relatively small 
cross sections, in the presence of a large background. 
The first step in this direction is
to measure a number of structure functions 
of the neutral kaon electroproduction cross section, as 
proposed by Magahiz {\it et al.} \cite{magahiz}.

\begin{table}[t]
\begin{center}
\begin{tabular}{cc}
{\epsfxsize=7cm \epsffile{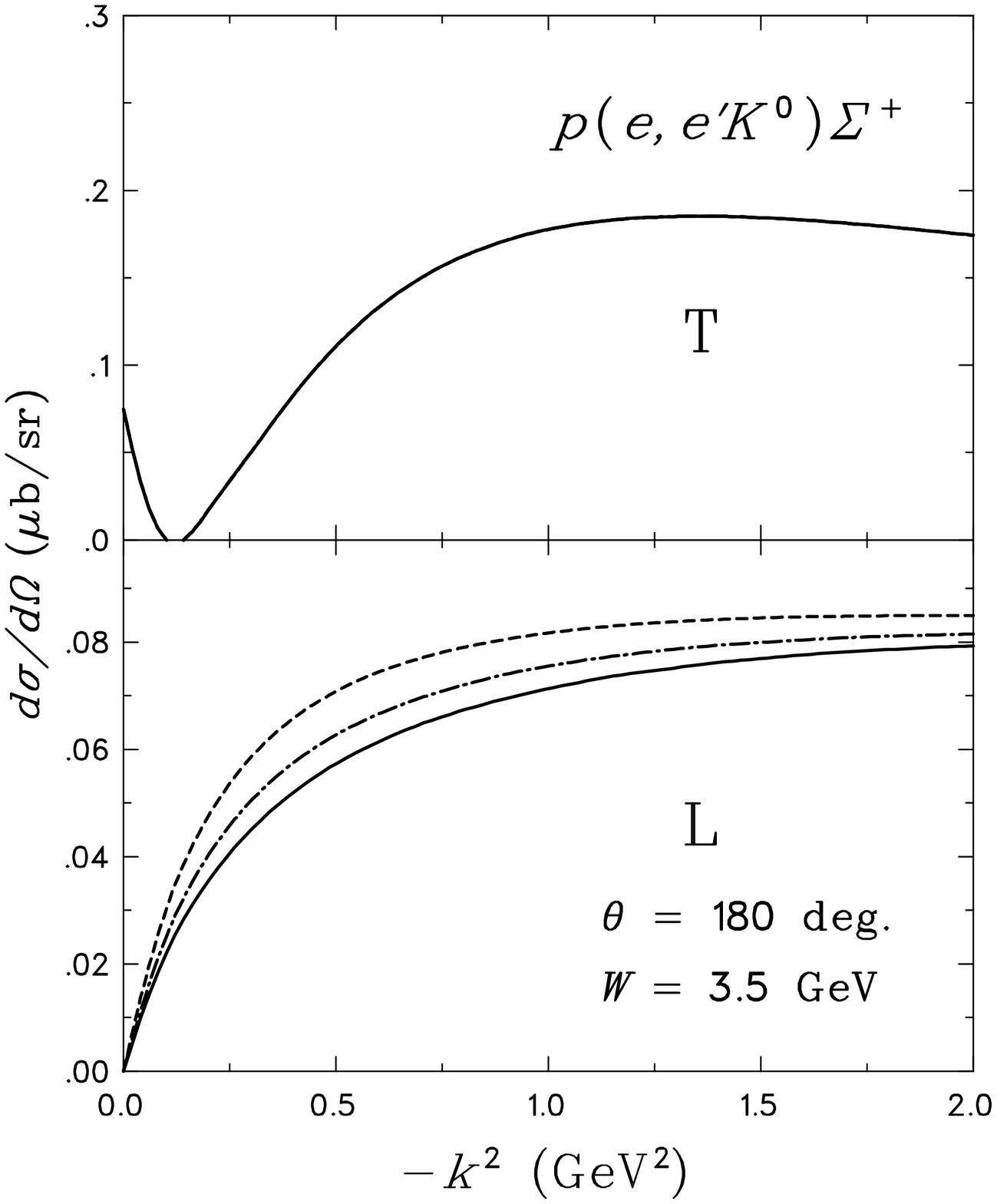}} &
{\epsfxsize=7cm \epsffile{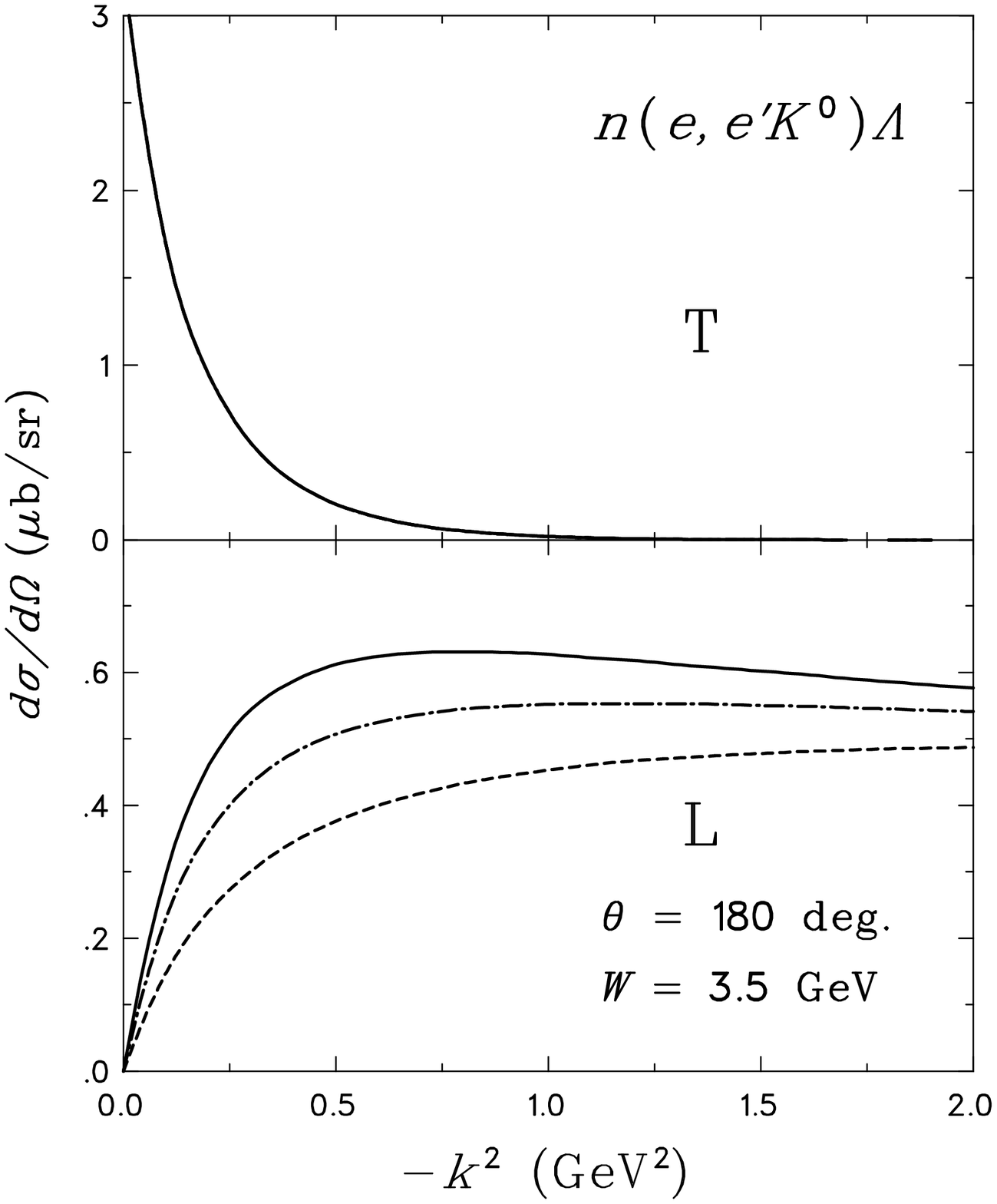}}\\
\end{tabular} 
\fcaption{Longitudinal and transverse cross sections for the reaction 
$p(e,e^{\prime}K^0)\Sigma^+$ (left) and
$n(e,e^{\prime}K^0)\Lambda$ (right). The calculations are with the neutral kaon
form factor predicted by the LCQ model (solid line) and the QMV model 
(dash-dotted line). The dashed line is obtained with no form factor.}
\label{fig:k0ff}
\end{center}
\end{table}

We have examined the sensitivity of the three reactions, 
$p(e,e^{\prime}K^0)\Sigma ^{+}, n(e,e^{\prime }K^0)\Sigma ^0$, 
and $n(e,e^{\prime }K^0)\Lambda $ to two different relativistic quark
models~\cite{ito1} of the $K^0$ form factor, the
light-cone quark (LCQ) and quark-meson vertex (QMV) model.  We point out
that the $K^0$ form factor is a prime candidate to be obtained from
lattice gauge calculations and therefore should be measured with high
priority.
As shown in Fig. 3, we found only very small effects in the observables for the
$\Sigma$ production reactions (see Fig.~\ref{fig:k0ff}).
This can be understood from the fact that the leading Born coupling
constant, $g_{K\Sigma N}$, which multiplies the $K^0$ $t$-channel
pole term, ranges between $0.1<|g_{K\Sigma N}/\sqrt{4\pi }|<1.0$. It is 
therefore much smaller than $g_{K\Lambda N}$, which governs the $K\Lambda$ 
Born terms and has a range of $2.0<|g_{K\Lambda N}/\sqrt{4\pi}|<4.4$.
Future experiments will therefore have to use the
$n(e,e^{\prime }K^0)\Lambda$ reaction, where we found sizable 
effects from the $K^0$ form factor.

As expected, the transverse cross section is not sensitive
to the $K^0$ form factor, because the
$K^0$ pole amplitude is proportional to the function of $k^2$ and vanishes 
at the photon point, $k^2=0$.
Therefore, the $K^0$ form factor contributes 
mostly to the L and LT cross sections. 
At backward angles, the predictions based on the various models are 
clearly distinguishable. The longitudinal cross 
section computed with the form factor of the LCQ-model is 
almost 50$\%$ larger than the calculation without $K^{0}$
pole terms, while the QMV-model calculation lies between those two.
While we only show the L and T structure functions here,
the sensitivity of the LT term is similar to the L term.
Thus, experiments at TJNAF's Hall B would not need to perform an
L/T separation in order to access the form factor.

In order to reduce the model dependency and to obtain accurate 
quantitative predictions, sufficient information from
kaon photoproduction experiments \cite{schu} is required. The
analysis of those measurements has to determine the relevant
resonances and coupling constants and thus define the
production amplitude as precisely as possible. 

\vspace*{0.8cm}
\noindent {\bf Acknowledgements} 

The work of CB has been supported in part by US Department of Energy under
contract no. DE-FG02-95-ER40907. TM and DK are supported in part by the
University Research for Graduate Education (URGE) grant.

\end{document}